\documentstyle[aps,prl,twocolumn]{revtex}

\begin{document}

\draft

\title{
Scaling Relations and Exponents in the Growth \\
of Rough Interfaces Through Random Media
}

\author{J.M. L\'{o}pez,$^{(1)}$ M.A.
Rodr\'{\i}guez,$^{(1)}$ A. Hern\'{a}ndez-Machado, $^{(2)}$ and
A. D\'{\i}az-Guilera $^{(3)}$}

\address{
$^{(1)}$Departamento de F\'{\i}sica Moderna, Universidad de
Cantabria,
and Instituto
de Estudios Avanzados en F\'{\i}sica Moderna y Biolog\'{\i}a
Molecular,
CSIC-UC,
Avenida Los Castros, E-39005 Santander, Spain. \\
$^{(2)}$Departament d'Estructura i Constituents de la Mat\`{e}ria,
Universitat de Barcelona,
Diagonal 647, E-08028 Barcelona, Spain. \\
$^{(3)}$Departament de F\'{\i}sica Fonamental,
Universitat de Barcelona,
Diagonal 647, E-08028 Barcelona, Spain.
}

\maketitle

\begin{abstract}

     The growth of a rough interface through a random media is
modelled by a continuous stochastic equation with a quenched
noise. By use of the Novikov theorem we can transform the dependence of
the noise on the interface height into an effective temporal
correlation for different regimes of the evolution of the
interface. The exponents characterizing the roughness of the
interface can thus be computed by simple scaling arguments
showing a good agreement with recent experiments and numerical
simulations.
\end{abstract}

\narrowtext

     The growth of rough interfaces through a random medium is
an interesting pattern-formation phenomena that has received a
lot of attention in recent years
\cite{r1,r2,r3,r4,r6,r7,r8,r9,r10,r11,r12,r13}.
In particular, experimental studies of the displacement of a
nonwetting fluid by a more
viscous wetting one through a porous media have shown that
the interface separating both
fluids is rough and the random pinning of the displacing fluid
on the wider porous of the medium strongly influences the
growth \cite{r2,r6}.
The random media is frozen in time and a natural way to
model its effects is by introducing a quenched disorder in a
continuous equation, as has been done by Kessler, Levine and
Tu (KLT)
\cite{r3}, Parisi \cite{r8},
and others \cite{r7,r11}.  This model is
different from the one that describes the growth of solids from
a vapor or from a stochastic ballistic deposition,
in which the stochastic field changes both in space and time.
These last phenomena are usually described by equations of the
type introduced by Edwards and Wilkinson (EW)
\cite{pie.ew}
and by Kardar, Parisi and Zhang (KPZ) \cite{prl56.889}. The
effects of
a quenched noise in a stochastic model are much less known and
are the subject of this Letter.

     We study the model introduced by KLT \cite{r3} and
Parisi \cite{r8}.  Starting from an initially flat
interface above a d-dimensional substrate at time $t=0$, a rough
interface is described by its height $h(\vec{x},t)$ at position
$\vec{x}$ and time $t$, which obeys the following stochastic
equation:
\begin{equation}
\frac{ \partial }{ \partial t}h(\vec{x},t)= \nabla ^{2}
h(\vec{x},t) +
F +  \eta (\vec{x},h),
\label{stochastic}
\end{equation}
where the diffusive term models the surface tension effects, $F$
is the pushing force and the noise $ \eta $ is Gaussian, with
zero mean value and delta correlated,
$$< \eta (\vec{x},y) \eta
(\vec{x'},y')>=\frac{ \theta }{2} \delta^{d} (\vec{x}-\vec{x'})
\delta (y - y'),$$
with an intensity $\theta$. This equation also models the growth
of domain walls in a random-field Ising ferromagnet \cite{r13}.

     Analytical results for the exponents associated with the
model described by Eq. (\ref{stochastic}) are only known in the
limit of strong pushing, where the interface moves very fast
(for $d=1$ the roughening exponent $ \alpha =1/2$ \cite{r3} and the
time exponent $ \beta =1/4$
\cite{r8}). In this limit, the model reduces to the case
of a noise with a delta correlation in time and the exponents
correspond to the EW model, for which the scaling relation
between exponents is given by $ \alpha / \beta =2$. On the other hand
for small pushing ($F<F_c$) the interface becomes pinned.

     KLT \cite{r3} and Parisi \cite{r8} have
performed numerical simulations of the model given by
Eq.(\ref{stochastic}) for different values of the pushing force.
Far from $ F_{c}$, Parisi \cite{r8} has obtained that,
after a transient, there is a temporal regime with anomalous
exponents followed by a crossover to an asymptotic regime that
corresponds to the EW model. As one approaches the pinning
transition, the crossover appears at later times, but the
long-time behavior is described in any case by the EW model.  Only at
the pinning transition one gets a different exponent, $
\beta=3/4$ \cite{r8}.
In the intermediate temporal regime, the numerical results of
KLT \cite{r3} give $ \alpha =0.73$. These results should
be compared with the experimental results of Rubio {\em et al.}
\cite{r2} ($ \alpha =0.73 \pm 0.03$) and Horvath
{\em et al.} \cite{r6} ($ \beta =0.65$ and $\alpha \sim 0.81$).

     The purpose of this Letter is to obtain analytical
expressions for the exponents characterizing the interface
growth and the associated scaling relations in regimes where
scaling properties are obeyed. To do so, we start with Eq.
(\ref{stochastic}) and we derive an effective model valid for
small intensities of the noise, $ \theta$. As we will see below,
the resulting effective model is given by the following
stochastic linear differential equation
\begin{equation}
\frac {\partial y}{\partial t} =  \nabla^2 y +
\xi(\vec{x},t)
\label{heigthfluc}
\end{equation}
where $y(\vec{x},t) = h(\vec{x},t) - \overline h(t)$ measures
the fluctuations of the height interface $ h(\vec{x},t)$ around
its averaged value $\overline h(t)$.  $\xi(\vec{x},t)$ is the
effective noise with zero mean value and correlation given by:
\begin{equation}
<\xi(\vec{x},t) \xi(\vec{x'},t')>= \frac{\theta}{2}
\delta^{d}(\vec{x}-\vec{x'})
W(t-t').
\label{corrnoise}
\end{equation}
The noise is delta-correlated in the substrate position and the
dependence on $h$ is substituted by a temporal correlation
$W(t-t')$. The explicit form of $W(t-t')$ will be discussed in
the following paragraph.

     From Eqs. (\ref{heigthfluc},\ref{corrnoise}) it is possible
to derive expressions for the width of the interface,
$\sigma=<y^{2}>^{1\over 2}$ and the correlation
lenght $ l_{c} $ in terms of
$W(t-t')$. Therefore, the scaling properties of $W(t-t')$ will
determine the roughening exponents and the associated scaling
relations.

\paragraph*{Derivation of the effective model.-} In order to get
an effective model,
we use functional methods (a generalization \cite{novikov} of the
Novikov theorem \cite{original}) to determine
the mean value of $\eta$:
$$ <\eta(\vec{x},h(\vec{x},t))> =
\frac {\theta}{2} \int_{0}^{t} G(0,t-t') dt' \times
$$
\begin{equation}
\times
\int dy \int dy'
\delta '(y-y')
 P_{\vec{x},t;\vec{x},t'}(y,y') + O(\theta ^2)
\label{disorder1}
\end{equation}
and also its correlation
$$
<\eta(\vec{x},h(\vec{x},t)) \eta(\vec{x}',h(\vec{x}',t'))> =
\frac {\theta}{2} \delta^{d}(\vec{x}-\vec{x}') \times
$$
\begin{equation}
\times
\int dy P_{\vec{x},t;\vec{x},t'}(y,y) + O(\theta^2)
\label{disorder2}
\end{equation}
where $G(\vec{x},t)\sim t^{-d/2} exp({\frac{-x^2}{t}})$
is the diffusive propagator and
\begin{equation}
P_{\vec{x},t;\vec{x}',t'}(y,y') =
<\delta (y - h(\vec{x},t))\delta (y' - h(\vec{x}',t'))>
\end{equation}
is the joint probability for the interface to have heights $y$
and $y'$ at $\vec{x},t$ and $\vec{x}',t'$ respectively. We note
that Eqs.(\ref{disorder1}) and (\ref{disorder2})
are valid for small values of the intensity of the noise because
we have neglected terms of order $\theta^2$ involving the joint
probability of four points.
By assuming homogeneity both in
space and time and defining $W(u,t-t')$ as the probability of
having an increment $u$ in a time $t-t'$, $W(u,t-t') =
P_{\vec{x},t;\vec{x},t'}(u+h,h) /P_{\vec{x},t'}(h)$, we obtain
the effective model given by
Eqs.(\ref{heigthfluc},\ref{corrnoise}) where
$\xi(\vec{x},t)=\eta(\vec{x},h(\vec{x},t)) -
<\eta(\vec{x},h(\vec{x},t))>$ is the effective noise with zero
mean value and correlation given by Eq.(\ref{disorder2}) up to
the lowest order in $\theta$, and $W(t-t') = W(0,t-t')$ is the
probability for the interface to remain at the same point after
a time interval $(t-t')$. Now, since the model is linear, the
kinetics of the interface is completely characterized by this
probability.

\paragraph*{Exponents and scaling relations.-}
First, we will discuss the different relevant length scales
associated with the interfacial growth. The interplay between
these scales will determine the different scaling regimes.  One
scale is given by the horizontal correlation length, $l_c$,
which for a very general form of $W(t-t')$, has diffusive
behavior with time, $l_c \sim t^{1/2}$. This result is
remarkable because it indicates a very universal behavior for $l_c$
in the complete temporal regime independent from the properties
of the disorder. It only depends on the diffusive character of
Eq.(\ref{heigthfluc}) and on the delta spatial correlation of
the noise. Another relevant scale is associated to the
characteristic length, $L_c$, of the clusters of points of the
interface that become pinned by the effect of the quenched
disorder. This cluster length depends on the pushing force. At
the pinning transition, with a pushing force $F_c$, the entire
interface becomes pinned and the cluster length is $L_c \sim L$.
By increasing $F$, $L_c$ is reduced and it becomes zero in the
limit of strong pushing, where the interface moves very fast and
the model reduces to EW model. In this limit, $W(t-t')$ becomes
a delta function. In accordance with these results, we assume a
scaling behavior of the form $L_c \sim (F-F_c)^{-\eta}$ where
$\eta$ is the associated exponent.
Starting from a flat interface at $t=0$ with a fixed
value of the pushing force $F$, the horizontal correlation
length, $l_c$, grows with time whereas the cluster length $
L_c$, remains at a fixed value. At some time, that we define as
the crossover time $t_c$, we have $l_c(t_c)= L_c$, and for times
$t>t_c$ the evolution is described by the EW model. From its
definition, $t_c$ has the following scaling relation, $ t_c \sim
(F-F_c)^{-\nu}$, where $\nu=2\eta$.

     Regarding the short time behavior for which $l_c < L_c$,
there is a regime with scaling properties in which the exponents
and the scaling relations could be defined and they would be
different from the ones associated with the EW model.  To
determine these quantities, we only need to realize that
$W(t-t')$ depends on the relevant scale in this regime, $L_c$.
However, due to the scaling properties, this is equivalent to
assuming a dependence on $L$.  In this way, we make a very general
assumption about the form that $W(t-t')$ scales with $L$ and $t$,
$W(t-t') \sim L^{-\Omega}\sigma ^{-\gamma}$ where $\Omega$ and
$\gamma$ are two independent exponents.  Now, substituting this
expression of $W(t-t')$ in the effective model, we obtain
a scaling dependence of $\sigma$ in $L$ even before the
saturation time is approached. Hence in this case the interface
width scales with time and size before it saturates ($t<t_s$)
as $\sigma \sim L^{\alpha '} t^{\beta}$ instead of the usual
scaling given by $\sigma \sim t^{\beta}$. The expressions for the exponent
$\beta$, the new exponent $\alpha'$ and the exponent $z$
associated with $l_{c}$ as $l_{c}\sim t^{z}$
are now given in terms of the parameters
$\Omega$ and $\gamma$ by:
\begin{equation}
\alpha '=-\frac{\Omega}{2+\gamma} \:\:\:\:
\beta   = \frac{4-d}{2(2+\gamma)} \:\:\:\:
z=2
\label{exp1}
\end{equation}

     From Eq.(\ref{exp1}), we can discuss the expression of
the $\alpha$ exponents that one gets for small system sizes for
which the saturation happens before the effects of the pinning
clusters have disappeared. The roughening exponent $\alpha$ is
calculated by taking $t=t_s \sim L^2$ in the expression for
$\sigma(t)$. The result is:
\begin{equation}
\alpha = \frac{4-d}{2+\gamma}- \frac{\Omega}{2+\gamma}
\label{exp2}
\end{equation}

\paragraph*{A model for temporal correlation $W(t-t')$.-}
In the previous paragraph, we characterized the two different
scaling
regimes. Now, we would like to
propose a simple model for $W(u,t-t')$ which
match these two regimes.  Based on physical arguments,
one could assume that $W(u,t-t')$ is a regular function peaked
on the mean height $u=\overline{h}(t-t')$ with a width $\sigma$.
As a simple assumption, we consider a lorentzian function.  In
the short time behavior it has to have a scaling dependence in
$L$ and $\sigma$. This could simply be done by means of the
following model
\begin{equation}
W(u,t) = \frac {\sigma}{\sigma^{2} + (\overline h
-u)^{2}}
\left( L^{\Omega} \frac {\sigma^{1+\gamma}}{\overline
h^{2}} + 1
\right) ^{-1}
\label{model}
\end{equation}
where $\gamma$ and $\Omega$ are the parameters that
characterized the model. The two different scaling regimes are
defined by $L^{\Omega} \gg \overline h^{2} \sigma^{-1-\gamma}$
and $L^{\Omega} \ll \overline h^{2} \sigma^{-1-\gamma}$. From
Eq.  (\ref{model}) we obtain that for long time, $ t\gg t_c$
the scaling behavior of $W(0,t)$ is $W(0,\lambda t) \sim \lambda
^{-1} \delta (t)$, as $\lambda \to \infty$, and $W(t-t')$
reduces to a delta function, as expected.  By substituting
$W(t-t')$ in the effective model one could get the exponents in
the different scaling regimes. The advantage in assuming an
expression for $W(t-t')$ over the complete temporal regime is
that one can determine the crossover time, $t_c$, that can be
defined by
$L^{\Omega} =
\overline h^{2}(t_c)\sigma^{-\gamma-1}(t_c)$.
By substituting $\overline h \sim t$ and the scaling behavior of
$\sigma$ one obtains
\begin{equation}
t_{c}\sim L^{\Omega \bigl(2-\beta(1+\gamma)\bigl)^{-1}(2+\gamma)^{-1}}
\label{crossover}
\end{equation}

\paragraph*{Numerical results and comparison.-}
At this point, all our exponents and scaling relations
associated with the first scaling regime depend on the
parameters $\Omega$ and $\gamma$. To corroborate the presence of
both scaling regimes and the values of these parameters, we have
performed numerical simulations of Eq.(\ref{stochastic}) for
$d=1$ with the same parameters as in Ref. \cite{r3}.
 In our numerical results, the pinning transition occurs at
$ F_c \sim 0.13 $.  We observe a first scaling regime with
$\beta \sim 3/4 $ in accordance with Parisi results. This result
implies that our parameter $\gamma$ is zero in this regime.
After a crossover $t > t_c$, $\beta \sim 1/4 $ like the EW
model.  In Fig. 1 we present our numerical results for
$l_c$ versus time in logaritmic scales. We observe the same
diffusive behavior for the complete temporal regime in
accordance with our model.

     Regarding the roughening exponent $\alpha$, our results are
in agreement with those obtained by different methods,
$\alpha =0.7-0.9$
for $t_s<t_c$ \cite{r2,r3,r4,r6,r12} and
$\alpha =0.49$ for $t_s>t_c$ \cite{r6}.
Furthermore, in the first regime, the
results $\beta=3/4$ and
$\alpha=0.8$ \cite{overhangs}
are compatible in our analytical
model by taking for our free
parameters the values $\gamma=0$ and $\Omega=1.4$.  Finally, in
Fig. 2 we present our numerical results for the crossover time
$t_{c}$ versus $F-F_c$. We obtain an exponent $\nu \sim 1.43$,
which in terms of $L$ is equivalent to $t_{c} \sim L^{2-\nu} =
L^{0.57}$.  This numerical result is in good agreement with our
analytical results obtained by substituting in
Eq.(\ref{crossover}) the values of $\Omega$, $\beta$ and
$\gamma$, i.e. $t_{c} \sim L^{0.56}$.

     In conclusion, our
analytical results predict a growth for the width in agreement
with different simulations and experiments with the appropriate
choice of two free parameters .
A scaling relation determines
the crossover time exponent. In our model, the correlation length
shows a diffusive behavior and we believe it has to be
analyzed in other models, like directed percolation (DP) and the
random
field Ising model, in order to clarify the diversity of growth
behaviors and hence universality classes.
For instance, in DP \cite{r10}
anisotropy generates
KPZ exponents after the crossover instead of those of EW.

     The authors wish to acknowledge H. Guo for very fruitful
discussions. This work has been supported by DGICyT of the
Spanish Government, projects \#PS90-0098 (J.M.L. and M.A.R.),
\#PB90-0030 (A.H.-M.), and \#PB92-0863 (A.D.-G.),
and NATO Collaborative Research Grant CRG.931018.

\begin{figure}
\caption{Horizontal correlation length versus time for two values
of $F$, close $F=0.15$ and far $F=0.4$ from the pinning
transition $F_c=0.13$.  The slope of the adjusted curve is
$1/2$.}
\end{figure}
\begin{figure}
\caption{Crossover time versus $F-F_c$ in log-log scale.
The adjusted curve gives an exponent $\nu=1.43$.}
\end{figure}


\begin{references}
\bibitem{overhangs}
The range of values for $\alpha$ obtained in different works
seems to
be due to a different interpretation of overhangs as is
discussed in \cite{r12}. Our model (\ref{stochastic}) does not contain
overhangs because the partial differential equation
has a single-valued solution.
\end{references}
\end{document}